\documentclass[preprint]{aastex6}
\usepackage{graphicx}
\usepackage{epsfig}
\usepackage{multirow}
\usepackage{footnote}

\shorttitle{{\it IUE} Solar Analog Observations}
\shortauthors{Lubin, Melis \& Tytler}
\begin{document}


\title{\large \bf Ultraviolet Flux Decrease Under a Grand Minimum from {\it IUE} Short Wavelength Observation of Solar Analogs}

%% Use \author, \affil, plus the \and command to format author and affiliation 
%% information.  If done correctly the peer review system will be able to
%% automatically put the author and affiliation information from the manuscript
%% and save the corresponding author the trouble of entering it by hand.
%%
%% The \affil should be used to document primary affiliations and the
%% \altaffil should be used for secondary affiliations, titles, or email.

%% Authors with the same affiliation can be grouped in a single
\author{\large Dan Lubin\altaffilmark{1}, Carl Melis\altaffilmark{2}, and David Tytler\altaffilmark{2}}
\affil{email: dlubin@ucsd.edu \\
$^1$Scripps Institution of Oceanography, University of California, San Diego, 9500 Gilman Drive, La Jolla, CA 92093-0221, USA \\
$^2$Center for Astrophysics and Space Sciences, University of California, San Diego, CA 92093-0424, USA \\
}

%\altaffiltext{1}{\centering Center for Astrophysics and Space Sciences, University of California, San Diego, CA 92093-0424, USA}
%\altaffiltext{2}{D\'{e}partement de Physique, Universit\'{e} de Montr\'{e}al, Montr\'{e}al, QC H3C 3J7, Canada}

%\affil{American Astronomical Society \\
%2000 Florida Ave., NW, Suite 300 \\
%Washington, DC 20009-1231, USA}

%\author{Butler Burton\altaffilmark{3}}
%\affil{National Radio Astronomy Observatory}

%\affil{TeXnology Inc}

%\author{Julie Steffen\altaffilmark{4}}
%\affil{American Astronomical Society \\
%2000 Florida Ave., NW, Suite 300 \\
%Washington, DC 20009-1231, USA}

%% Use the \and command so offset the last author.
%\and

%\author{Jeff Lewandowski\altaffilmark{5}}
%\affil{IOP Publishing, Washington, DC 20005}


%\altaffiltext{1}{AAS Journals Data Scientist}
%\altaffiltext{2}{greg.schwarz@aas.org}
%\altaffiltext{3}{AAS Journals Associate Editor-in-Chief}
%\altaffiltext{4}{AAS Director of Publishing}
%\altaffiltext{5}{IOP Senior Publisher for the AAS Journals}

\begin{abstract}

\large{
We have identified a sample of 33 Sun-like stars observed by the {\it International Ultraviolet Explorer (IUE)} with the short wavelength (SW) spectrographs that have ground-based detections of chromospheric Ca\,II H+K activity. Our objective is to determine if these observations can provide an estimate of the decrease in ultraviolet (UV) surface flux associated with a transition from a normal stellar cycle to a grand minimum state. The activity detections, corrected to solar metallicity, span the range $-$5.16 $<$ log $R'$$_{HK}$ $<$ $-$4.26, and eight stars have log $R'$$_{HK}$ $<$ $-$5.00. The {\it IUE}-observed flux spectra are integrated over the wavelength range 1250$-$1910 \AA , transformed to surface fluxes, and then normalized to solar B $-$ V. These normalized surface fluxes show a strong linear relationship with activity $R'$$_{HK}$ ($R^2$ $=$ 0.857 after three outliers are omitted). From this linear regression we estimate a range in UV flux of 9.3\% over solar cycle 22, and a reduction of 6.9\% below solar cycle minimum under a grand minimum. The 95\% confidence interval in this grand minimum estimate is 5.5\%$-$8.4\%. An alternative estimate is provided by the {\it IUE} observations of $\tau$\,Cet (HD\,10700), a star having strong evidence of being in a grand-minimum state, and this star's normalized surface flux is 23.0$\pm$5.7\% lower than solar cycle minimum.
}

\end{abstract}




\large{

\section{\large \bf Introduction}

Over the past decade there has been increasing realization and concern that the steady and high solar luminosity of the past century may transition to greater variability later this century (Abreu et al.\ 2008; Feulner \& Rahmstorf 2010; Lockwood 2010). Specifically, the Sun may descend into a period of low magnetic activity analogous to the historical Maunder minimum (MM, circa 1640$-$1715; Eddy 1976). A resulting decrease in total solar irradiance (TSI) impacting the terrestrial lower atmosphere energy budget is linked to changes in high-latitude circulation patterns that strongly influence the climate of Europe and the Atlantic sector of the Arctic and subarctic (Song et al.\ 2010; Meehl et al.\ 2013), and may also influence Antarctic climate (Orsi et al.\ 2012). Studies have also shown the importance of stratospheric response to a grand minimum (e.g., Gray et al.\ 2010; Bolduc et al.\ 2015; 
Maycock et al.\ 2015). Over a solar cycle and certainly in response to a future grand minimum, irradiance variability at middle ultraviolet (UV) wavelengths that drive oxygen photolysis and ozone chemistry is much larger that that of the TSI. Resulting changes to stratospheric ozone abundance alter the stratosphere-troposphere temperature gradient and feed back to tropospheric planetary wave refraction, further altering climatically relevant circulation patterns (Maycock et al.\ 2015). With this realization that both direct radiative and indirect stratospheric influences affect terrestrial climate under a solar grand minimum, it is important to understand how UV irradiance would respond to such a large and prolonged change in solar magnetic activity.

      Spectral UV irradiance during the historical MM is usually inferred from time series reconstructions grounded by the distribution of solar surface features over the course of current cycles or by geophysical proxies (e.g., Lean 2000; Krivova et al.\ 2010; Shapiro et al.\ 2011; Bolduc et al.\ 2014). Solar analogs have mainly been employed to make individual stellar comparisons with the Sun (e.g., Judge et al.\ 2014), or in surveys to estimate the percentage of very low activity stars occurring in the local field population as an indicator of how frequently grand minima occur throughout the Sun's lifetime (e.g., Baliunas \& Jastrow 1990; Henry et al.\ 1996; Hall \& Lockwood 2004; Lubin et al. 2010, 2012; Curtis et al.\ 2017). In this study we explore whether an aggregate sample of {\it International Ultraviolet Explorer (IUE)} observations of Sun-like stars, acquired for multiple unrelated observing programs between 1978 and 1996 and maintained as archival data, can be used to estimate the decrease in UV flux associated with a grand minimum.

\section{\large \bf Data}
\label{secobs}

Solar analogs were selected from the {\it IUE} MAST archive for which there is at least one observation of Ca\,II H+K chromospheric activity reported in the literature and at least one useful {\it IUE} short wavelength (SW) spectrum (most stars have multiple spectra). All IUE spectra considered here were obtained at low resolution ($\sim$6 \AA ). In individual SW spectra, an absolute calibration uncertainty of 5\% is an appropriate estimate (Bohlin 1986). However, in this work we must also consider dispersion between individual spectra obtained from stars with varying magnitudes. Table 1 of Bohlin et al.\ (1990) gives examples for {\it IUE} SW data. In these examples dispersions are of order 4-5\% for hot stars with strong UV flux, but become larger for nearby G stars (8.9\% for $\beta$\,Hyi) and even larger still (35-58\%) for fainter G stars. We therefore expect there might be large variance among spectra for some stars. To evaluate uncertainty in the averaged fluxes for each solar analog, we use the standard deviation of the wavelength-integrated fluxes for stars with multiple spectra. For stars with just one spectrum we adopt a standard deviation of 9\% based on Bohlin et al.\ (1990).

      The three criteria for selecting a solar analog are (1) location within $\pm$0.5\,mag of the {\it Hipparcos} main sequence (HMS; Wright 2004); (2) effective temperature $T_e$ from the Geneva-Copenhagen survey (Holmberg et al.\ 2009) within $\pm$500\,K of solar; and (3) surface gravity consistent with a dwarf star. For each SW spectrum, the measured flux was integrated over 1250$-$1910 \AA , a wavelength range containing the absorption spectrum of molecular oxygen (Schumann-Runge system; e.g., Wayne 1991). This wavelength-integrated flux was converted to a surface flux $F_s$ using the relation of Oranje et al.\ (1982) as described in Buccino \& Mauas (2008). Distances to these stars range from 3.7 to 32.2\,pc, so we make no correction for interstellar extinction. The sample of 33 stars and their properties are listed in Table 1, including the average $F_s$ for each star and its standard deviation used in this study. Table 1 also includes (for most stars) two independent observations of log $g$, and for all stars log $g$ $>$ 4.0.

      Observations of Ca\,II H+K flux for these stars are taken from Duncan et al.\ (1991), Baliunas et al.\ (1995), Henry et al.\ (1996), Strassmeier et al.\ (2000), Tinney et al.\ (2002), Gray et al.\ (2003), Wright et al.\ (2004), Jenkins et al.\ (2008), Isaacson \& Fischer (2010), Arriagada (2011) and Lovis et al.\ (2011). In all these reports except for Lovis et al.\ (2011) the Ca\,II H+K fluxes are given in Mt.\ Wilson S-units (Duncan at al.\ 1991). The S-units are transformed to log $R'$$_{HK}$ by removing the photospheric component using the procedure of Noyes et al.\ (1984), described in Henry et al.\ (1996) and Wright et al.\ (2004).

      Metallicity variation influences the photospheric opacity and introduces corresponding variability in Ca\,II H+K flux. The log $R'$$_{HK}$ values reported by Lovis et al.\ (2011) have already been corrected to solar metallicity. We adjust other all observations considered here to solar metallicity following Sousa et al.\ (2008). For this adjustment we preferred spectroscopic metallicity determinations when available. If metallicity observations were available in Cayrel de Strobel et al.\ (2001) these were averaged. Otherwise the metallicity was taken from Valenti \& Fischer (2005), or if not available there, then from the multi-band photometric estimate in the Holmberg et al.\ (2009).

      Metallicity also influences a star's displacement from the HMS. An evolved subgiant can lie on the HMS if it has sufficiently low metallicity. A subgiant has such a contrasting internal structure, age and rotational history from the Sun that we must take care not to accidentally include definitively evolved stars in our analysis (Jenkins et al.\ 2008; Saar \& Testa 2012). We used the parameterization of Twarog et al.\ (2009) to check that for any evolved stars that might appear close to the HMS due to significantly non-solar [Fe/H]. Using this correction we find only two stars (HIP\,33719 and HIP\,64792) with heights above the HMS of 0.504 and 0.527 magnitudes, respectively. Both are slightly metal-rich stars, and may be very slightly evolved, but we retained them in the analysis based on their log $g$.

      Figure 1 shows individual surface flux spectra from four solar analogs: (1) HIP\,8102 ($\tau$\,Cet), a cooler star than the Sun (B $-$ V $=$ 0.727) and with steady and low activity (average solar-corrected log $R'$$_{HK}$ $=$ $-$5.031) such that it is widely regarded as a possible grand-minimum star (e.g., Judge et al.\ 2004; Hall et al.\ 2009); (2) HIP\,57443, a star with near-solar temperature, activity slightly less than mean solar (B $-$ V $=$ 0.664 and log $R'$$_{HK}$ $=$ $-$4.973); (3) HIP\,15457, a slightly cooler star than the Sun (B $-$ V $=$ 0.681) but active (log $R'$$_{HK}$ $=$ $-$4.419); and (4) HIP\,1599, a hotter star than the Sun (B $-$ V $=$ 0.576) but with activity only slightly greater than mean solar (log $R'$$_{HK}$ $=$ $-$4.911). As expected, the surface flux (left panel of Figure 1) depends very strongly on temperature, with HIP\,1599 appearing much brighter than the others. Upon correcting the fluxes to solar temperature (discussed below), the same four objects show differences that correlate with activity. HIP\,15457 appears brighter than the others throughout most of the spectral range, and there is a small difference between the grand-minimum candidate HIP\,8102 and the solar-activity-range star HIP\,57443.

\section{\large \bf Analysis}
\label{secspec}

We see a strong relationship between surface flux and B $-$ V, similar to what Buccino \& Mauas (2008) find for the high-resolution data near the Mg\,II emission feature. Figure 2(a) shows all individual observations. More outliers appear in this plot than in Figure 2 of Buccino \& Mauas, and we reject these outliers on the assumption that they are poorly flux-calibrated observations. To uniformly apply this assumption we identify these outliers using the modified Thompson tau test, and continue iterating on all observations for a given star until either (1) no more outliers are identified, or (2) the standard deviation of the remaining observations is less than 5\%, whichever occurs first. Out of 211 {\it IUE} observations, the modified Thompson tau test rejects only eight. Four surface flux observations are significantly larger than the final mean for the star, and four are smaller. All of the stars from which these individual observations were rejected had ten or more observations, and no more than two observations were rejected per star. Therefore this initial outlier identification should not bias our final result. The average surface fluxes $F_s$ for each star after these outlier rejections are shown in Figure 2(b).

      Our objective is to determine the relationship between $F_s$ and the Ca\,II H+K chromospheric activity flux $F'$$_{HK}$ $=$ $R'$$_{HK}$$\sigma$$T_e$$^4$. Plotting these {\it IUE}-measured $F_s$ averages directly as a function of $F'$$_{HK}$ (Figure 3(a)) reveals only a weak relationship. To assess the true dependence on surface flux on activity these average $F_s$ must be normalized to solar temperature. We expect a nonlinear relationship between surface flux and B $-$ V, and we tried weighted linear regressions of both $F_s$$^{1/4}$ and log $F_s$ against B $-$ V (weighted linear regression being less sensitive to outliers than ordinary least squares). The regression using log $F_s$ gives a slightly better correlation. This regression yields slope $-$4.812, intercept 9.710, correlation $-$0.879 and $R^2$ $=$ 0.773. After using this regression to normalize the average fluxes to solar B $-$ V, we find a strong relationship between these normalized surface fluxes and activity $F'$$_{HK}$ (Figure 3(b)). In Figure 3(b) a weighted linear regression yields slope 1.2260, intercept 1.9989 $\times$ 10$^6$, correlation 0.859 and $R^2$ $=$ 0.738.

      There is still considerable scatter in Figure 3(b), particularly among the lowest activity stars. We therefore decided to omit any stars that appear as outliers in a standard statistical test. We apply the Iglewicz \& Hoaglin (1993) test for multiple outliers, which with a modified Z-score $\geq$3.5 flags three stars as possible outliers: HIP\,37853, HIP\,47592 and HIP\,51248. HIP\,37853 has the lowest metallicity of our sample ([Fe/H] $=$ $-$0.884), which might explain its $F_s$ lying well above the regression line. HIP\,47592 is the hottest star in our sample (B $-$ V $=$ 0.534) and, lying at the extremity of our B $-$ V regression, may not have been completely corrected to solar temperature. HIP\,51248 has only one {\it IUE} observation, and based on Figure 2(a) we might have less confidence on its $F_s$ than those of stars with multiple observations. The remaining 30 average surface fluxes normalized to solar B $-$ V are plotted versus $F'$$_{HK}$ in Figure 4(a). A weighted linear regression here yields a slope of 1.2130, intercept 2.0098 $\times$ 10$^6$, correlation 0.926 and $R^2$ $=$ 0.857. Thus the omission of the three outliers reduced uncertainty but did not significantly alter the slope or intercept of the resulting linear regression.

      Metallicity variation should generally affect the relationship between $F_s$ and $F'$$_{HK}$, so we tested the sensitivity in our derived slope and intercept by considering only the stars with metallicity closest to solar. Twenty stars in our sample have $-$0.1 $<$ [Fe/H] $<$ +0.2 (Figure 4(b)), and the regression with these stars (Figure 4(c)) yields a slope 1.1076, intercept 2.3560 $\times$ 10$^6$, correlation 0.884 and $R^2$ $=$ 0.781. This implies slightly less sensitivity in $F_s$ to $F'$$_{HK}$, but is not significantly different than our result in Figure 4(a).

      With the linear regression in Figure 4(a) we can estimate the reduction in surface flux under a grand minimum relative to a typical solar minimum. Woods \& Rottman (2002) used NASA {\it Upper Atmosphere Research Satellite (UARS)} observations to evaluate the maximum 
and minimum spectral solar surface fluxes for cycle 22 for wavelengths shorter than 2000 \AA . For the interval 1250$-$1910 \AA , the surface flux at cycle 22 minimum is 9.3\% smaller than at maximum. Baliunas et al.\ (1995) report Mt.\ Wilson Ca\,II H+K observations for the Sun over cycle 22, and from their Figure 1(d) we obtain S-unit values of 0.195 and 0.168 for cycle 22 maximum and minimum, respectively. Following Noyes et al.\ (1984), these correspond to log $R'$$_{HK}$ values of $-$4.822 and $-$4.951, respectively, for solar B $-$ V $=$ 0.656. Evaluating the linear regression at these values gives surface fluxes of 3.164 $\times$ 10$^6$ and 2.867 $\times$ 10$^6$\,erg\,cm$^{-2}$\,s$^{-1}$, respectively. This difference is 9.4\%, consistent with the Woods \& Rottman (2002) observations. Schr\"{o}der et al.\ (2012) suggest a quiet Sun value of S $=$ 0.15, based on observations of solar activity during cycle 24 minimum on five days in early 2009 when the solar disk was entirely plage-free. With S $=$ 0.15 Noyes et al.\ (1984) gives log $R'$$_{HK}$ $=$ $-$5.065 and our linear regression of Figure 4 gives a corresponding surface flux 2.669 $\times$ 10$^6$\,erg\,cm$^{-2}$\,s$^{-1}$. This is 6.9\% lower than our estimated value for cycle 22 minimum. The 95\% confidence intervals on the slope and intercept of this regression are 1.0210$-$1.4050 and 1.7271$-$2.2926 $\times$ 10$^6$, respectively. This corresponds to a confidence interval for differences in surface fluxes (a) between solar maximum and solar minimum of 7.7\%$-$11.2\%, and (b) between solar minimum and quiet Sun of 5.5\%$-$8.4\%. These confidence limits assume that the scatter around the regression line in Figure 4 among the inactive stars is mainly due to flux calibration uncertainty. If we allow the possibility that the scatter represents real stellar variability (probably true for some of the stars), then a wider 95\% confidence interval of approximately $\pm$33\% applies to any star in the solar activity or inactive range, based on standard confidence estimation for the predicted value in a linear regression model. At present we cannot resolve which confidence limits apply, not having suitable time series observations of both the activity $R'$$_{HK}$ and the UV surface flux $F_s$ for all of the stars.

\section{\large \bf Discussion}
\label{secdisc}

Reconstructions for UV irradiance from the historical MM to the present day vary widely. Krivova et al.\ (2010) estimate a difference between the MM and recent solar minima of 5.1\% in the Schumann-Runge continuum (1300$-$1750 \AA ) and 1.9\% in the Schumann-Runge bands (1750$-$2000 \AA ). Lean (2000) reconstructs a difference between MM and mean present-day solar spectral irradiance of $\sim$15\% increasing to $\sim$30\% as wavelength decreases from $\sim$1900 \AA\ to $\sim$1300 \AA . Our estimate for the interval 1250$-$1910 \AA\ is 15.6\% between quiet Sun and solar maximum, 11.1\% between quiet Sun and median present-day solar activity, and 6.9\% between quiet Sun and cycle 22 minimum. Our solar analog estimate therefore appears to be intermediate between the Lean (2000) and Krivova et al.\ (2010) reconstructions.

      In contrast, Shapiro et al.\ (2011) estimate a difference of 26.6\% between the MM and recent solar minima in the Schuman-Runge bands (1750$-$2000 \AA ). The only way we can construe such a large difference in our {\it IUE} SW sample is if we consider HIP\,8102 ($\tau$\,Cet), which is the one star in our sample regarded as a likely grand minimum analog. We estimate its mean activity from the literature (Table 1, corrected to solar metallicity) at log $R'$$_{HK}$ $=$ $-$5.031. We have eight SW observations of HIP\,8102, none of which were flagged as outliers, and which give an average 1250$-$1910 \AA\ surface flux (normalized to solar B $-$ V) of 2.207 $\times$ 10$^6$\,erg\,cm$^{-2}$\,s$^{-1}$ with a standard deviation of 5.7\%. This is 23.0\% lower than cycle 22 minimum, and the second lowest value in our sample. It is not clear that we should assert such a large difference based on just one star in our relatively small sample, given that we have no comparable information about the cycling states of most of our other stars in the very low activity or solar activity ranges. Judge et al.\ (2004) provide high-resolution spectroscopic evidence in the 1306$-$1551 \AA\ range that $\tau$\,Cet may be in a genuine grand minimum phase. At the same time, $\tau$\,Cet is metal poor and so has a shallower convection zone than the Sun for its mass and $T_e$, and this might influence its activity.

      Limitations with this study include the small sample size, unknown quality of the flux calibration, and possibly incomplete metallicity correction of log $R'$$_{HK}$ values from the literature. Although published {\it IUE} SW calibration gives state-of-the-art uncertainties for observations of standard stars early in the {\it IUE} mission (Bohlin 1986), the scatter in our selected observations led us to reject a few of them as poorly flux-calibrated data. With an early space telescope operating continuously for nearly 18 years, one can naturally expect that not every observation would realize ideal instrument performance. For example, it is possible that some observations did not center the object properly on the aperture. There is a possibility that our somewhat conservative outlier identification procedure for individual spectra did not flag all of the poorly flux-calibrated data (three averages in Figures 2(b) and 3 have large standard deviations: HIP\,37583, HIP\,64394, HIP\,96901). We note that the solar $F_s$ values on our Figure 4(a) regression line are $\sim$20\% larger than the {\it UARS} measurements (Table 1) reported by Woods \& Rottman (2002), although our difference between cycle 22 maximum and minimum is highly consistent with the {\it UARS} measurements. This is probably a manifestation of the overall flux calibration uncertainty in the {\it IUE} data. Another limitation with the small sample size is an inability to find many observations that are contemporaneous with the ground-based log $R'$$_{HK}$ detections. This introduces uncertainty in the activity values of the most inactive stars, although this uncertainty is probably smaller than those related to the flux calibration. Finally, examination of the metallicity correction in Lovis et al.\ (2011) and Sousa et al.\ (2008) reveals that the correction applies only to the bolometric flux term and does not fully correct the $R'$$_{HK}$ calibration. Based on the metallicity distribution in our sample (Figure 4(b)), we expect this limitation to have minimal influence on our result although it should be considered when working with a larger sample in future work. Despite these limitations, we find it noteworthy that archival {\it IUE} data acquired between two and four decades ago for a variety of scientific programs can be compiled here to make a defensible estimate of the change in UV flux between a recent solar cycle minimum and a grand minimum.

\acknowledgments

This work was supported by California state funds to the Scripps Institution of Oceanography. All of the data presented in this paper were obtained from the Mikulski Archive for Space Telescopes (MAST). STScI is operated by the Association of Universities for Research in Astronomy, Inc., under NASA contract NAS5-26555. Support for MAST for non-HST data is provided by the NASA Office of Space Science via grant NNX09AF08G and by other grants and contracts. We thank the anonymous referee for several insightful suggestions that improved the analysis and manuscript.

}



\clearpage

\begin{figure}
 \centering
 %\hspace{-0.3in} 
 \includegraphics[width=200mm]{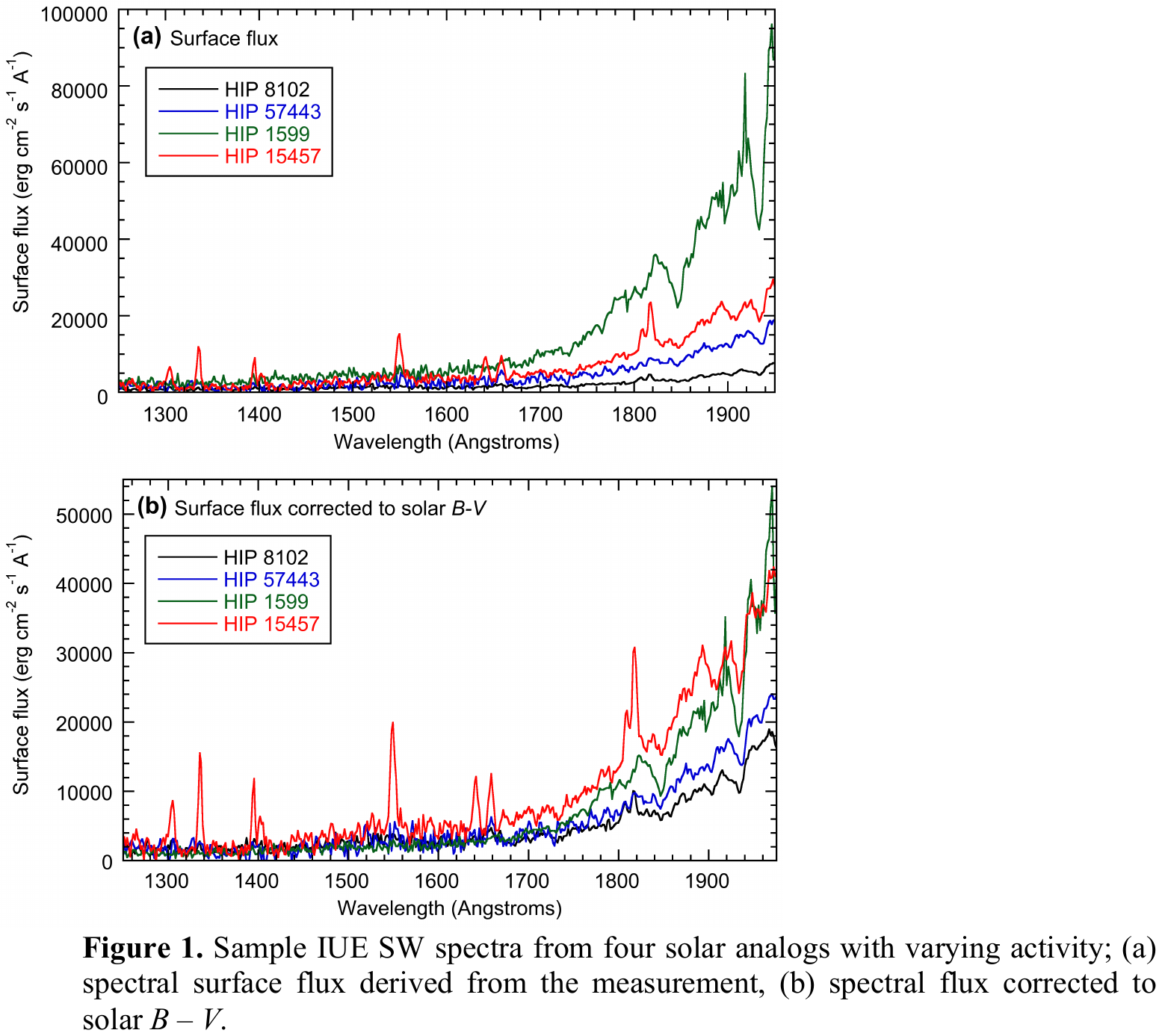}
\end{figure}

\clearpage

\begin{figure}
 \centering
 \includegraphics[width=200mm]{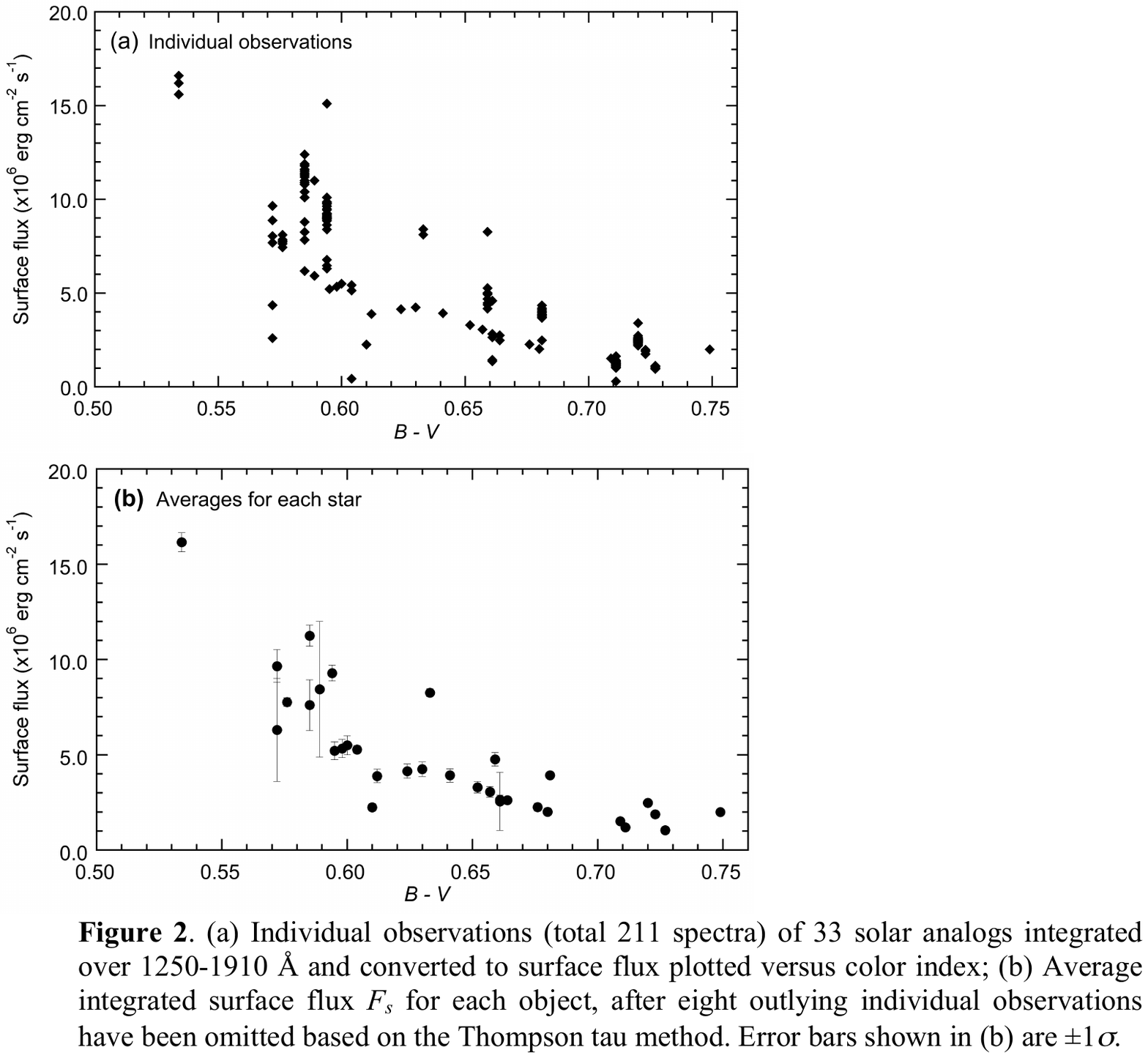}
\end{figure}

\clearpage

\begin{figure}
 \centering
 \includegraphics[width=200mm]{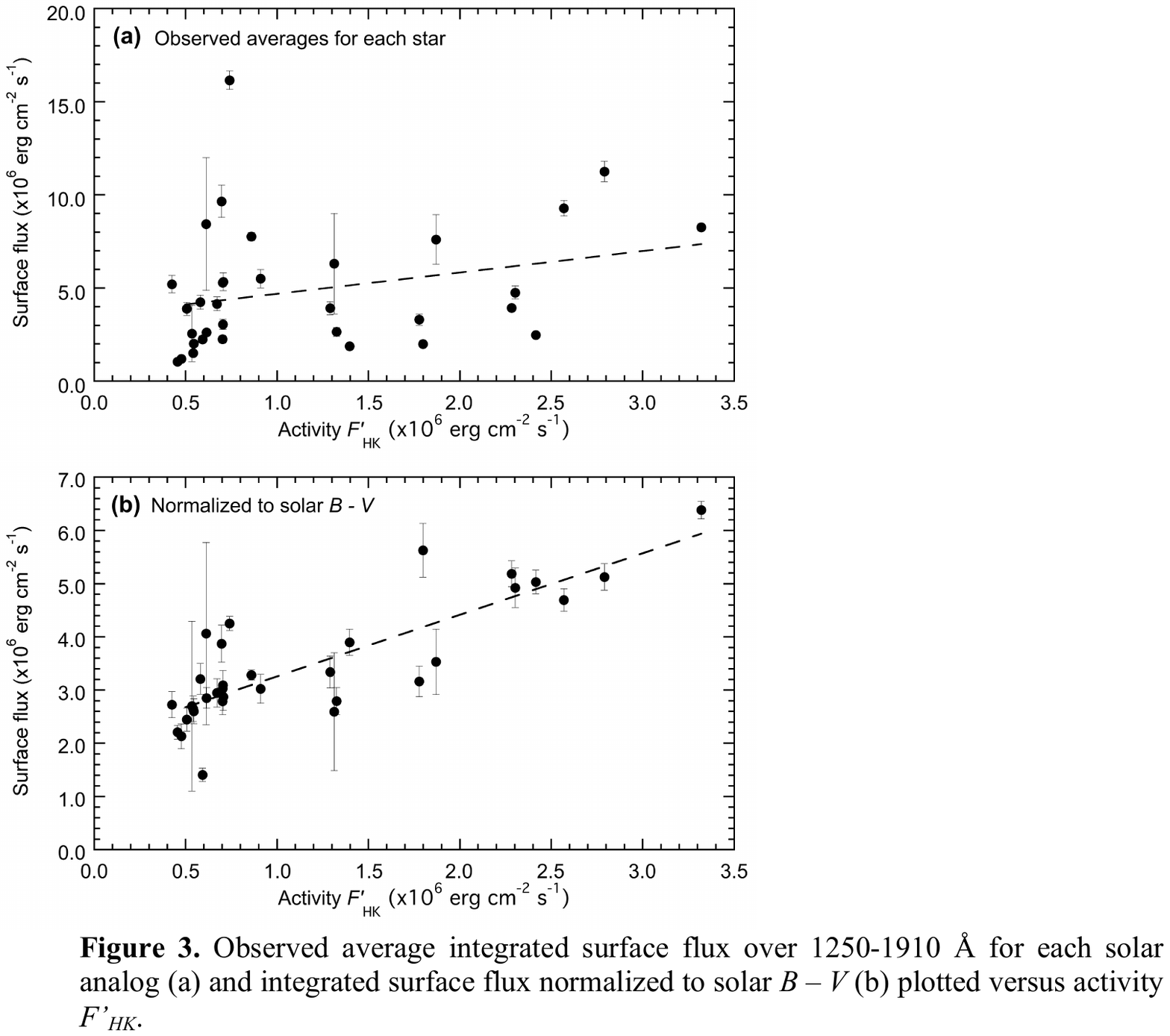}
\end{figure}

\clearpage

\begin{figure}
 \centering
 \includegraphics[width=200mm]{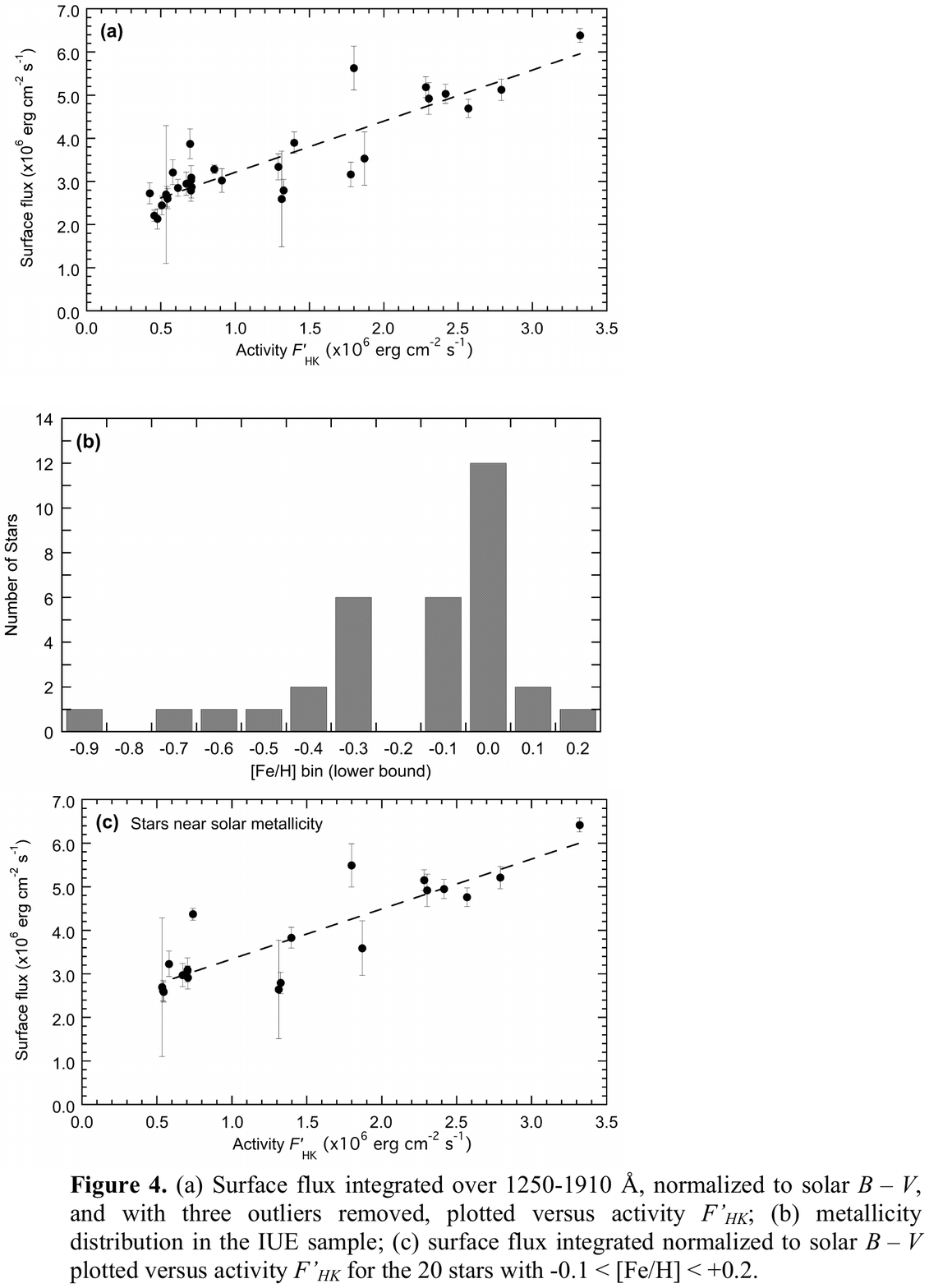}
\end{figure}

\clearpage

%\floattable

\begin{deluxetable}{lccccccccc}
%\rotate
\tabletypesize{\small}
\tablecolumns{10}
\tablewidth{0pt}
\tablecaption{\large{{\it IUE} SW Solar Analogs}}
\tablehead{ 
  \colhead{HIP} & 
  \colhead{HD} & 
  \colhead{$B$ $-$ $V$} & 
  \colhead{$T_e$ (K)} & 
  \colhead{$<$[Fe/H]$>$} & 
  \colhead{$<$log $R'$$_{HK}$$>$} & 
  \colhead{$R'$$_{HK}$ source} & 
  \colhead{$F_s$} & 
  \colhead{$\sigma$($F_s$)} & 
  \colhead{log $g$}
}
\startdata
            &     Sun    &    0.656     & 5777   &  0.000  &   $-$4.822    &     2     &     2.462      &          &        \\
            &     Sun    &    0.656     & 5777   &  0.000  &   $-$4.951    &     2     &     2.234      &          &        \\
1599    & 1581      & 0.576       & 5929   & $-$0.232 & $-$4.911    &   3, 11  &    7.763      & 0.223 & 4.51(1), 4.46(2) \\
1803    & 1835      & 0.659        & 5689   &   0.177    & $-$4.412   & 1$-$3,7   & 4.762     & 0.357 & 4.44(1), 4.47(2) \\
4290    & 5294      & 0.652        & 5728    & $-$0.210 & $-$4.536 & 1, 9         &  3.301     & 0.297  & 4.48(3) \\
8102     & 10700    & 0.727       & 5420     & $-$0.539 & $-$5.031 & 1$-$3, 5, 7, 9$-$11   & 1.049  & 0.060  & 4.31(1), 4.59(2) \\
15330   & 20766   & 0.641       & 5702      & $-$0.208   & $-$4.667  & 3, 5  & 3.923   & 0.353   & 4.39(1), 4.58(2) \\
15371     &    20807   &  0.600    & 5835   & $-$0.206  & $-$4.858   & 3, 11  & 5.504   & 0.495   & 4.43(1), 4.54(2) \\
15457     & 20630      & 0.681    & 5702   & 0.007  & $-$4.419   & 1, 2, 9     & 3.931    & 0.182   & 4.40(1), 4.49(2) \\
15510    & 20794       & 0.711    & 5458   & $-$0.394  & $-$5.024  & 3, 11   &  1.200    & 0.132 & 4.30(1), 4.62(2) \\
24813    & 34411       & 0.630    & 5875   & $-$0.002   & $-$5.066  & 1, 6, 7, 9  & 4.243   & 0.382   & 4.16(1), 4.37(2) \\
27913    & 39587       & 0.594     & 5902   & $-$0.025   & $-$4.428  &  1, 6, 9  & 9.288  &  0.422   & 4.39(1), 4.34(2) \\
30104   & 44594       & 0.657     & 5754    & 0.130     & $-$4.946    & 3, 11      & 3.057  & 0.275   & 4.43(1), 4.41(2) \\
33719   & 52265       & 0.572     &  6081   & 0.210    & $-$5.046    &  7, 9       &  9.655  & 0.869   & 4.29(1), 4.26(2) \\
37853    & 63077     & 0.589     & 5794    & $-$0.884   & $-$5.019  & 3, 7       & 8.438   & 3.559   & 4.01(1) \\
42291     & 73524    & 0.598    & 5902    & 0.075    & $-$4.989    & 3, 11       & 5.335    & 0.480   & 4.32(1), 4.41(2) \\
43726    & 76151      & 0.661    &  5676   & 0.034    & $-$4.647   & 1$-$3     & 2.649    & 0.238   & 4.41(1), 4.55(2) \\
44897    & 78366      & 0.585    & 5916   & 0.080    & $-$4.570    & 1, 2, 6, 9   & 7.607  & 1.327   & 4.54(2) \\
47592    & 84117      & 0.534    & 6138    & $-$0.070   & $-$5.036   & 9    & 16.156     & 0.501   &  4.31(2) \\
51248   & 90508       & 0.610   & 5728    & $-$0.330   & $-$5.013   & 1, 6  & 2.247    & 0.202    &  4.35(1) \\
53721   & 95128       &  0.624   & 5834   & $-$0.014    & $-$4.990   & 1, 7   & 4.149  & 0.374   & 4.24(1), 4.38(2) \\
56997  & 101501     & 0.723     & 5483   & 0.030       & $-$4.565     & 2, 6, 7   & 1.881  & 0.119  & 4.69(1), 4.43(2) \\
57443   & 102365    & 0.664    & 5649   & $-$0.330   & $-$4.973     & 3, 9       & 2.619   & 0.179   & 4.32(1), 4.57(2) \\
64394   & 114710    & 0.572    & 5970   & 0.014      & $-$4.739      & 1, 2, 7, 9  & 6.311  & 2.697  & 4.34(1), 4.57(2) \\
64792 & 115383     & 0.585     & 5957   & 0.075      & $-$4.408     & 1$-$3, 7, 9 & 11.253  & 0.551  & 4.11(1), 4.60(2) \\
64924  & 115617    &  0.709    & 5572   & $-$0.028  & $-$5.004    & 1$-$3, 7, 9, 11   & 1.509   & 0.136 & 4.27(1), 4.47(2) \\
72659  & 131156    &  0.720    & 5570   & 0.020      & $-$4.348    & 1$-$3, 6, 7   & 2.479  & 0.110  & 4.60(1), 4.65(2) \\
75676  & 138004     & 0.676    & 5888    & $-$0.470  & $-$4.900  & 6, 7, 9  & 2.256   & 0.203   & 4.49(2) \\
77257  & 141004     & 0.604    & 5888     & $-$0.022   & $-$4.987  & 1$-$3, 6, 7, 9   & 5.285   & 0.199 & 4.11(1), 4.30(2) \\
78459  & 143761     & 0.612    & 5781     & $-$0.248   & $-$5.097   & 1, 2, 6, 7, 9     & 3.892   & 0.350 & 4.14(1), 4.36(2) \\
82588  & 152391     & 0.749    & 5420     & 0.040        & $-$4.435    & 1$-$3, 6, 7    & 1.996    & 0.180 & 4.57(2) \\
85042   & 157347    & 0.680     & 5623    & 0.007       & $-$5.017    & 6, 7, 9, 11      & 2.018     & 0.182 & 4.36(1), 4.50(2) \\
88945    & 166435    & 0.633    & 5728     & 0.040     & $-$4.265     & 6, 7, 9            & 8.264     & 0.211 & 4.44(2) \\
96185 &  184499     & 0.595    &  5728     & $-$0.634  & $-$5.157   & 1                   & 5.210     & 0.469  & 4.01(1), 4.11(3) \\
96901  & 186427     & 0.661    & 5702      & 0.053        & $-$5.050   & 1, 6, 7, 9      & 2.557      & 1.515  & 4.37(1), 4.35(2) \\
\enddata
\tablecomments{(a) Surface fluxes $F_s$ are $\times$10$^6$\,erg\,cm$^{-2}$\,s$^{-1}$. (b) Solar $F_s$ values are at cycle 22 maximum and minimum reported by Woods \& Rottman (2002). (c) Activity sources are (1) Duncan et al.\ (1991); (2) Baliunas et al.\ (1995); (3) Henry et al.\ (1996); (4) Strassmeier et al.\ (2000); (5) Tinney et al.\ (2002); (6) Gray et al.\ (2003); (7) Wright et al.\ (2004); (8) Jenkins et al.\ (2008); (9) Isaacson \& Fischer (2010); (10) Arriagada (2011); (11) Lovis et al.\ (2011); (d) log $g$ sources are (1) Cayrel de Strobel et al.\ (2001); (2) Valenti \& Fischer (2005); (3) Ram\'{i}rez et al.\ (2012).}
\end{deluxetable}

\end{document}